\documentclass[twocolumn,article,superscriptaddress,letterpaper]{revtex4}

\usepackage{graphicx}
\usepackage{dcolumn}
\usepackage{bm}
\usepackage{amsmath, amssymb}%
\usepackage{epstopdf}
\usepackage{color}
\usepackage{soul}
\usepackage{hhline}
\usepackage{multirow}
\usepackage[usestackEOL]{stackengine}
\usepackage{subfig}

\setstackgap{L}{\normalbaselineskip}

\usepackage{float}
\floatstyle{plaintop}
\restylefloat{table}

\newcommand{\prel}{\ifmmode p_{rel} \else $p_{rel}$ \fi}
\newcommand{\pcm}{\ifmmode p_{cm} \else $p_{cm}$ \fi}

\usepackage[]{changes} 

\begin{document}

\title{Comment on ``Searching for flavor dependence in nuclear quark behavior''}

\newcommand*{\MIT }{Massachusetts Institute of Technology, Cambridge, Massachusetts 02139, USA}
\newcommand*{\MITindex}{1}
\newcommand*{\ODU}{Old Dominion University, Norfolk, Virginia 23529}
\newcommand*{\ODUindex}{2}
\newcommand*{\JLAB}{Thomas Jefferson National Accelerator Facility, Newport News, Virginia 23606}
\newcommand*{\JLABindex}{3}
\newcommand*{\UW}{Department of Physics, University of Washington, Seattle, WA 98195}
\newcommand*{\UWindex}{4}
\newcommand*{\TAU }{School of Physics and Astronomy, Tel Aviv University, Tel Aviv 69978, Israel}
\newcommand*{\TAUindex}{5}
\newcommand*{\Penn}{Pennsylvania State University, University Park, PA, 16802}
\newcommand*{\Pennindex}{6}

\author{O. Hen}
\email[Contact Author \ ]{hen@mit.edu}
\affiliation{\MIT}
\author{F. Hauenstein}
\affiliation{\ODU}
\author{D.W. Higinbotham}
\affiliation{\JLAB}
\author{G.A. Miller}
\affiliation{\UW}
\author{E.~Piasetzky}
\affiliation{\TAU}
\author{A. Schmidt}
\affiliation{\MIT}
\author{E.P. Segarra}
\affiliation{\MIT}
\author{M. Strikman}
\affiliation{\Penn}
\author{L.B. Weinstein}
\affiliation{\ODU}

\date{\today{}}

\begin{abstract}

  \noindent Weinstein, et. al \cite{weinstein11} [PRL {\bf 106},
  052301 (2011)] and Hen, et. al \cite{hen12}
  [PRC {\bf 85}, 047301 (2012)] observed a correlation between the EMC effect and the
  amount of short range correlated (SRC) pairs in nuclei which implies that
  quark distributions are different in SRC pairs as compared with free
  nucleons.  Schmookler, et. al \cite{Schmookler:2019nvf} [Nature {\bf 566}, 354 (2019)] bolstered this by showing
  that the EMC data can be explained by a universal modification of
  the structure of nucleons in neutron-proton SRC pairs and presented
  the first data-driven extraction of this universal
  modification function (UMF).

  \noindent Arrington and Fomin \cite{Arrington:2019wky} [arxiv 1903.12535] attempt to gain
  insight into the correlation between the EMC effect and SRCs by
  distinguishing between correlated nucleon pairs at high-virtuality
  (HV) vs. high local-density (LD). However, there is an inconsistency
  in their derivations of the UMFs, $F_{univ}^{LD}$ and
  $F_{univ}^{HV}$, causing a non-physical difference between them for
  asymmetric nuclei. In addition, the combinatorial scaling they used
  to extract high-LD $np$, $pp$ and $nn$ pairs from measured HV $np$ pairs is
  contradicted by realistic ab-initio Quantum Monte-Carlo (QMC)
  calculations.

\end{abstract}

\maketitle


Ref.~\cite{Arrington:2019wky} attempts to study universal pair
modification functions (UMF)
for two cases: high-virtuality (HV) $np$-SRC pairs and SRC
pairs at high local density (LD).  High-LD $NN$ pairs are defined as having small
separation with either high or low relative momenta.  HV pairs have
both high momentum and small separation, making them a subset of high-LD
pairs.  

\section{Universal function derivation}
Eq. 1 in Ref.~\cite{Arrington:2019wky} was derived in~\cite{Schmookler:2019nvf} by modeling $F_2^A$ as:
\begin{eqnarray}
\begin{split}
F_2^A|_{HV} = & (Z-n_{np_{HV}}^A)F_2^p + (N-n_{np_{HV}}^A)F_2^n \\
&+ n_{np_{HV}}^A (F_2^{p*_{HV}} + F_2^{n*_{HV}}) \\
= & Z F_2^p  + N F_2^p + n_{np_{HV}}^A (\Delta F_2^{p_{HV}} + \Delta F_2^{n_{HV}}),
\label{Eq1}
\end{split}
\end{eqnarray}
where $n_{np_{HV}}^A$ is the number of HV $np$-SRC pairs in nucleus
$A$, $F_2^{p*_{HV}}$ and $F_2^{n*_{HV}}$ are the modified proton and
neutron structure functions, and $\Delta F_2^{p_{HV}} =F_2^{p*_{HV}}
- F_2^p$ (and similarly for the $\Delta F_2^{n_{HV}}$). All of the
functions $F$ depend on $x=Q^2/2m\omega$.  This assumes that almost
all HV (i.e., high momentum) nucleons belong to $np$-SRC pairs and
neglects the contribution of $nn$ and $pp$ pairs.  This approximation was shown
experimentally and theoretically to be good to better than 10\% \cite{Duer:2018sxh}.  The corresponding UMF is given by:
\begin{eqnarray}
\begin{split}
F_{univ}^{HV} &= n_{np_{HV}}^d \frac{\Delta F_2^{p_{HV}} + \Delta F_2^{n_{HV}}}{F_2^d}\\
&=\frac{\frac{F_2^A}{F_2^d}-(Z-N)\frac{F_2^p}{F_2^d}-N}{(A/2)a_2-N}
\label{Eq2}
\end{split}
\end{eqnarray}

Ref.~\cite{Arrington:2019wky} compares $F_{univ}^{HV}$ with what they
claim to be an equivalent expression for the high-LD assumption
$F_{univ}^{LD}$.  
Their function, Eq. 2 of
Ref.~\cite{Arrington:2019wky}, can be obtained by assuming the UMF is
related to the modified EMC-SRC correlation between the slope of
$R_{EMC}^A$ for $0.3\leq x_B\leq0.7$ and $R_2
\frac{A(A-1)}{2NZ}$~\cite{Arrington12}. However, there is no
theoretical justification for equating this expression with the
left-hand side of Eq.~\ref{Eq2}. Thus, it cannot be consistently
compared with $F_{univ}^{HV}$.

To consistently compare $F_{univ}^{HV}$ and $F_{univ}^{LD}$, Eq.~\ref{Eq1} needs to be re-written for high-LD pairs, which include $nn$ ($n_{nn_{LD}}^A$) and $pp$ ($n_{pp_{LD}}^A$) pairs:
\begin{eqnarray}
\begin{split}
&F_2^A|_{LD} =  (Z-n_{np_{LD}}^A-2n_{pp_{LD}}^A)F_2^p + \\
&~~~~~~~~~~~~~(N-n_{np_{LD}}^A-2n_{nn_{LD}}^A)F_2^n +\\
&n_{np_{LD}}^A (F_2^{p*_{LD}} + F_2^{n*_{LD}}) + 2n_{pp_{LD}}^A F_2^{p*_{LD}}  + 2n_{nn_{LD}}^A F_2^{n*_{LD}} \\
&~~~~~~~~~= Z F_2^p  + N F_2^p + \\
&~~~~~~~~~n_{np_{LD}}^A [ (1+\frac{Z-1}{N}) \Delta F_2^{p_LD} + (1+\frac{N-1}{Z}) \Delta F_2^{n_{LD}}],
\label{Eq3}
\end{split}
\end{eqnarray}
where  $n_{pp_{LD}}^A = n_{np_{LD}}^A  \frac{Z(Z-1)}{2NZ}$,
$n_{nn_{LD}}^A = n_{np_{LD}}^A
\frac{N(N-1)}{2NZ}$~\cite{Arrington:2019wky}.  

As $\Delta F_2^{p_{LD}}$ and $\Delta F_2^{n_{LD}}$ have different nucleus-dependent coefficients, unless one assumes a constant relation between them, Eq.~\ref{Eq3} cannot be used to extract an equivalent UMF to $F_{univ}^{HV}$ (i.e. equation that has the same left-hand side as Eq.~\ref{Eq2}) .

If instead we assume symmetric nuclei ($N = Z$) we get Eq. 2 of Ref.~\cite{Arrington:2019wky}:
\begin{eqnarray}
\begin{split}
F_{univ}^{LD} = n_{np_{LD}}^d \frac{\Delta F_2^{p_{LD}} + \Delta F_2^{n_{LD}}}{F_2^d}
=\frac{\frac{F_2^A/A}{F_2^d/2}-1}{\frac{A(A-1)}{2NZ}R_2-1}.
\label{Eq4}
\end{split}
\end{eqnarray}

Therefore, the difference between $F_{univ}^{HV}$ and $F_{univ}^{LD}$
comes primarily from the use of Eq.~\ref{Eq4} (that is only comparable
to $F_{univ}^{HV}$ for {\it symmetric} nuclei) for {\it asymmetric}
nuclei. This is done by defining the isoscalar-corrected EMC ratio $R_{EMC}^A=\frac{F_2^A/A}{F_2^d/2}/C_{isospin}$ and assuming that $R_{EMC}^A$ for asymmetric nuclei equals $\frac{F_2^A/A}{F_2^d/2}$ for a symmetric nucleus with the same $A$ but $Z = A / 2$.

This assumption is unjustified, especially if the EMC effect in
asymmetric nuclei is flavor-dependent. Moreover, it is not
consistently applied to $F_{univ}^{HV}$, which leads to an artificial
difference between $F_{univ}^{HV}$ and $F_{univ}^{LD}$.  This difference is largely
driven by the $(Z-N)\frac{F_2^p}{F_2^d}$ term that was 
artifically removed from $F_{univ}^{LD}$ but not from
$F_{univ}^{HV}$. This is inconsistent with the flavor dependence being
studied and casts doubt on the entire HV – LD comparison of
Ref.~\cite{Arrington:2019wky}.

\begin{figure} [t]
\includegraphics[width=0.9\columnwidth, height=5.5cm]{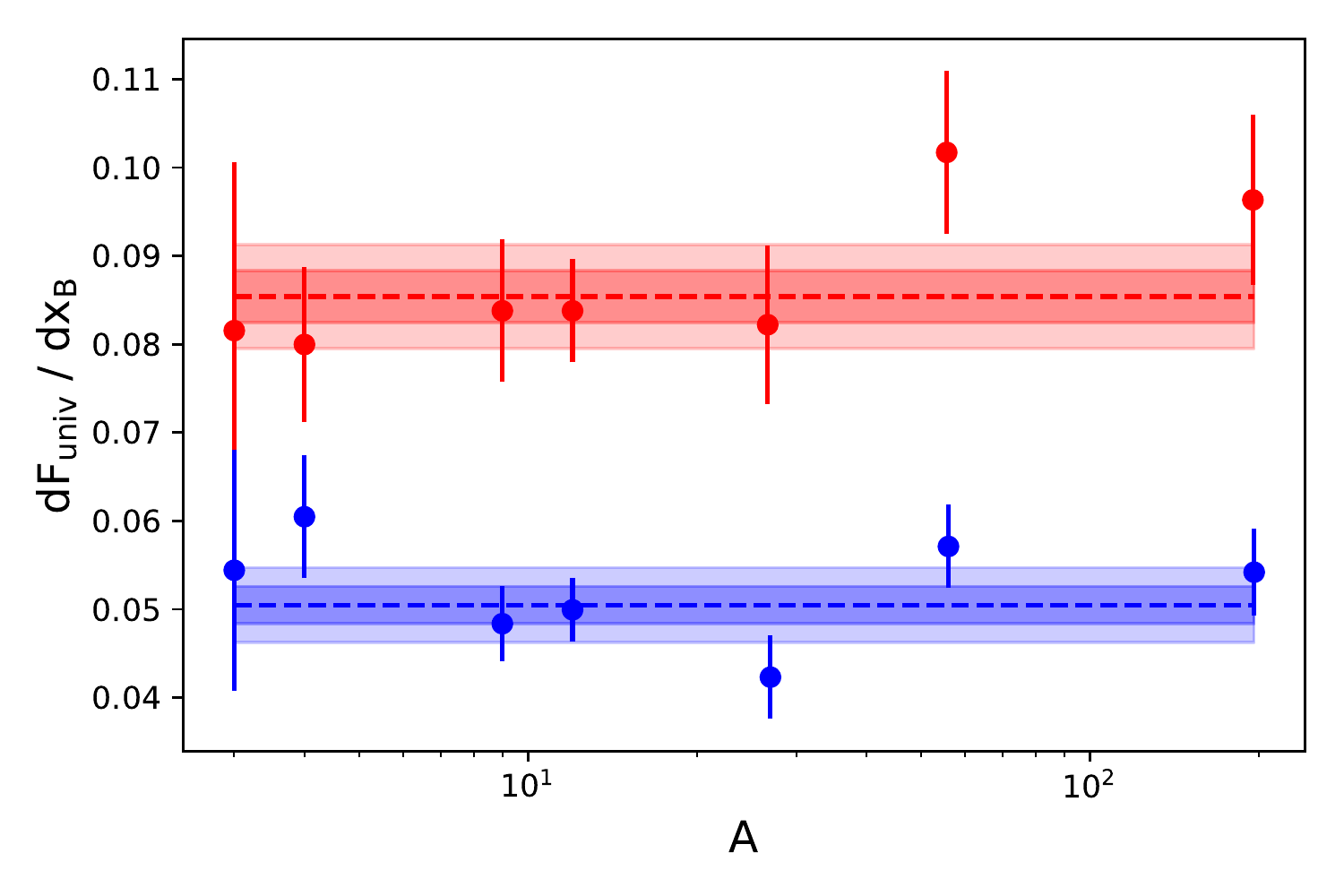}
\caption{One-parameter fits                                                                                                                    
of the UMF slopes $dF_{univ}^{HV}/dx$  (red) and $dF_{univ}^{LD}/dx$ from 
Eq.~\ref{Eq4} (blue) versus atomic number $A$.
The shaded bands indicate
the 68\% and 95\% confidence intervals.
Though these constant fits fully describe both UMF slopes,
in Fig. 3 of Ref.~\cite{Arrington:2019wky} 
the authors add yet another fit parameter leading to 
over-fitting the data, see text for details.}
\label{Fig:fit}
\end{figure}

In addition, Arrington and Fomin logarithmically fit the
$A$-dependence of the slopes of $F_{univ}^{HV}$ and $F_{univ}^{LD}$ (Fig.~3
\cite{2010arXiv1012.3754A}) in order to show which one is
more consistent with $A$-independence.  However, they failed to note that a
one-parameter constant fit to $dF_{univ}^{HV}/dx$ already gives a $\chi^2$/dof of $0.83$
and their two-parameter over-fitting gives $\chi^2$/dof of $0.34$
(Fig.~\ref{Fig:fit}).  For $F_{univ}^{LD}$ constant and logarithmic
fits give reduced $\chi^2$/dof of $1.3$ and $1.5$ respectively, again
indicating that a constant fit is more appropriate~\cite{2010arXiv1012.3754A}.

\begin{figure} [t]
\includegraphics[width=0.9\columnwidth, height=5.5cm]{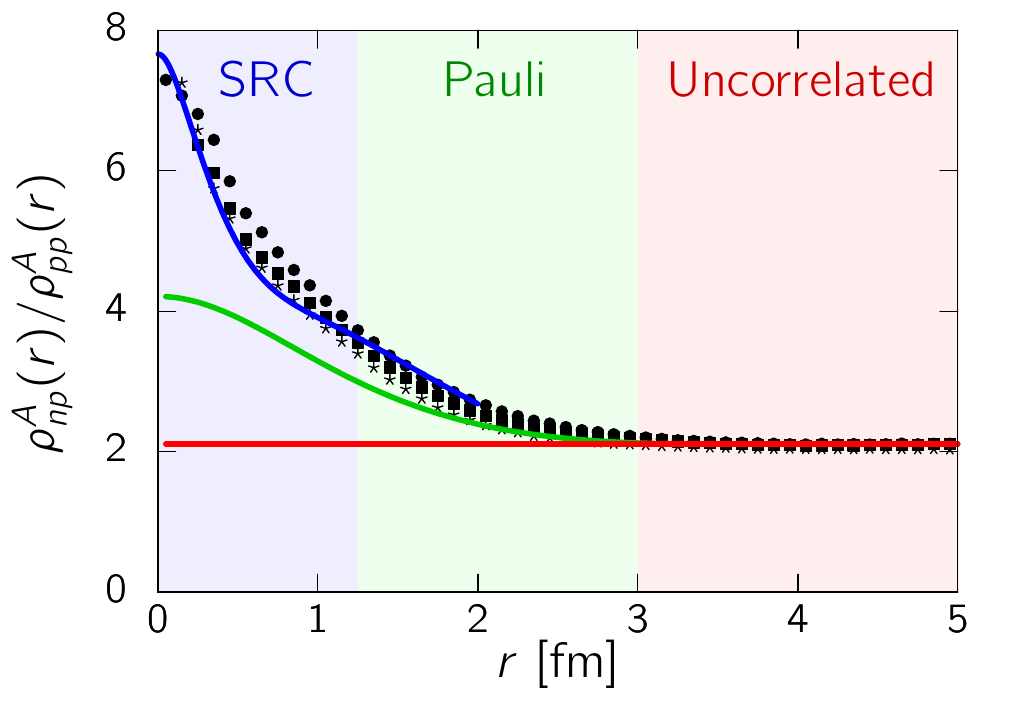}
\caption{QMC calculated ratio of $np$ / $pp$ pair density as a function of pair separation (points) for $^{12}$C (circles), $^{16}$O (squares) and $^{40}$Ca (stars)~\cite{wiringa14, Lonardoni:2017egu}, compared with calculations~\cite{Cruz-Torres:2017sjy} of SRC-dominated densities (blue) vs. uncorrelated pairs with (green) and without (red) accounting for Pauli exclusion.}
\label{Fig:combinatorics}
\end{figure}

\section{Combinatorial  scaling}
The combinatorial relations assumed for $n_{np_{LD}}^A$,
$n_{nn_{LD}}^A$ and $n_{pp_{LD}}^A$ are also questionable. LD and HV
correspond to different dynamical pictures as HV pairs are
predominantly $D$-wave while high-LD have increased $S$-wave contributions.

Ref.~\cite{Arrington:2019wky} assumes the high-LD ratio of $np$ pairs
to $pp$ pairs, $\rho_{np}^A(r)/\rho_{pp}^A(r)$, equals
$NZ/[Z(Z-1)/2]\approx2$ for symmetric nuclei.  However, ab-initio
calculations for $^{12}$C, $^{16}$O and $^{40}$Ca~\cite{wiringa14,
  Lonardoni:2017egu} indicate that this ratio is only 2 at large-r ($> 3$ fm), but
increases at small-r ($< 1$ fm) by a factor of 2 to 4 (see
Fig.~\ref{Fig:combinatorics}), implying a much smaller $pp$ and $nn$ pair
contribution at small-r.

Therefore the contribution of high-LD $pp$ and $nn$ pairs should be
reduced from the simplistic combinatorial calculation by a factor of 2
to 4, reducing the difference between $F_{univ}^{HV}$ and
$F_{univ}^{LD}$ from a factor of 2.5 to about 1.5.  Furthermore, in
the spin-0 channel, even for asymmetric nuclei, calculations show the
same abundances of small-$r$ $nn$, $pp$, and $np$ pairs, contrary to
combinatorial expectations~\cite{Weiss:2016obx}.

Recent work~\cite{Lynn:2019vwp} even showed that $NN$-pair scaling
coefficients at small-$r$ are the same for a remarkable range of $NN$
potentials (i.e, they are scale- and scheme-independent) and
consistent with measured values of $a_2(A/d)$ without requiring any
combinatorial scaling.

\section{Pair c.m. motion corrections}
Ref.~\cite{Arrington:2019wky} also distingush between scaling of HV
and high-LD pairs by defining seperate scale factors: $a_2$ (HV) and
$R_2$ (high-LD). The two are related by a multiplicative factor
arrising from the center of mass (c.m.) motion of SRC pairs.

c.m. motion effects can increase the measured $(e,e’)$ cross-section
ratio that is used to extract $a_2$. Correcting for this enhancement
is reasonable. However, it should be applied in the extraction of the
relative number of either high-LD or HV pairs and requires detailed
modeling of the nuclear spectral
function~\cite{Ciofi07,Weiss:2018tbu}. The application of this
correction only for the high-LD case, again, leads to an artificial
difference between the two approaches.

Quantitatively, Ref.~\cite{Arrington12} estimated the c.m. correction
to be 20\% for medium and heavy nuclei, using a simplistic
one-dimensional smearing of the deuteron momentum distribution. This
procedure ignored the three-dimensional nature of the problem and,
most importantly, the phase-space correlations that significantly
affect the measured electron scattering cross section. A more detailed
study~\cite{vanhalst12}, accounting for these and other effects,
suggested a 70\% correction factor.

\section{HV vs. LD scaling}
QMC calculations extract pair distributions in both coordinate
($\rho_{NN,\alpha}^A(r)$) and momentum space
($n_{NN,\alpha}^A(k)$). These densities were shown to both factorize
as~\cite{Weiss:2015mba,Weiss:2016obx,Cruz-Torres:2017sjy,Weiss:2018tbu,ryckebusch15,CiofidegliAtti:2017xtx,CiofidegliAtti:2017tnm,Alvioli:2016wwp,Alvioli:2013qyz,neff15}:
\begin{eqnarray}
\begin{split}
n_{NN,\alpha}^A (k>k_F ) = C_{NN,\alpha}^A \times |\psi_{NN,\alpha}(k)|^2 , \\
\rho_{NN,\alpha}^A (r<1 fm) = C_{NN,\alpha}^A \times |\psi_{NN,\alpha}(r)|^2 ,
\label{Eq5}
\end{split}
\end{eqnarray}
where $\alpha$ marks the pair spin-isospin state and
$\psi_{NN,\alpha}$ are zero-energy solutions of the two-body
Schrodinger equation for state $\alpha$. Their $k$- and $r$-space
representations are related by a Fourier transform that does not
change their normalization. $C_{NN,\alpha}^A$ are nucleus-dependent
scale factors that (A) account for the many-body dynamics and (B) are
the same in both $k$- and $r$-space for all spin-isospin channels.  This
single scaling factor at both small distance and large momentum is
inconsistent with the Ref.~\cite{Arrington:2019wky}  concept of small-$r$, low-$k$ correlated
pairs.

Eq.~\ref{Eq5} was
shown~\cite{Weiss:2016obx,Cruz-Torres:2017sjy,Weiss:2018tbu} to
reproduce QMC calculations at high-$k$ and small-$r$ to ~10\% for $A$ = 4 –
40 nuclei and describes electron-scattering data using the same
scaling factors $C_{NN,\alpha}^A$ as obtained from the QMC
calculations.

Therefore, ab-initio calculations do not support the existance of
different high-$k$ and small-$r$ scaling factors as used by
Ref.~\cite{Arrington:2019wky}, showin complete physical equivalence in
the many-body dynamics of HV and high-LD pairs.

\section{Conclusions}
The underlying cause of the EMC effect is an open question with far
reaching implications for our understanding of QCD effects in the
nuclear medium.  The original observations of the EMC-SRC
correlation~\cite{weinstein11, Hen12} and UMF
extraction~\cite{Schmookler:2019nvf}, raises an interesting and
relevant question about the mechanism driving this physics.

We explained that inclusive electron scattering data fundamentally cannot answer this question and pointed to a collection of quantitative issues with the analysis of Ref.~\cite{Arrington:2019wky}.

\begin{acknowledgments}
This work was supported by the U.S. Department of Energy, Office of Science, Office of Nuclear Physics under Award Numbers DE-FG02-97ER-41014, DE-FG02-94ER40818, DE-FG02-96ER-40960, and DE-AC05-06OR23177  under which Jefferson Science Associates operates the Thomas Jefferson National Accelerator Facility, the Pazy foundation, and the Israeli Science Foundation (Israel) under Grants Nos. 136/12 and 1334/16. 
\end{acknowledgments}

\bibliography{EMC_Barak_bib}

\end{document}